\documentclass{kluwer}
\usepackage{graphicx}
\begin{document}
\begin{article}
\begin{opening}
\title{Basic Properties of Compressible MHD Turbulence:\\ 
~~~~~~~~~~~~~~~~~~~Implications for Molecular Clouds 
}            
\author{A. Lazarian}
\author{J. Cho\email{lazarian, cho@astro.wisc.edu}}
\institute{University of Wisconsin-Madison, Dept. of Astronomy}


\runningtitle{Turbulent Molecular Clouds}
\runningauthor{Lazarian \& Cho}

\begin{abstract} 
Recent advances in understanding of the basic properties of compressible
Magnetohydrodynamic (MHD) turbulence call for revisions of some of the 
generally accepted concepts. First,  MHD turbulence is not so messy
as it is usually believed. In fact, the 
notion of strong non-linear coupling of compressible and 
incompressible motions is not tenable. Alfven, slow and fast modes of
MHD turbulence follow their own cascades and exhibit degrees of anisotropy
consistent with theoretical expectations. Second, the fast decay of turbulence
is not related to the compressibility of fluid. Rates of decay of 
compressible and incompressible motions are very similar. Third, 
 viscosity by neutrals does not suppress MHD turbulence
in a partially ionized gas. Instead,
 MHD turbulence develops magnetic cascade at scales below the scale at
which neutrals damp ordinary hydrodynamic motions. The implications of
those changes of MHD turbulence paradigm for molecular clouds  
require further studies. Those studies can
benefit from testing of theoretical predictions using 
new statistical techniques that utilize spectroscopic data.
We briefly discuss advances in development of 
tools using which the statistics of turbulent velocity can be recovered from
observations.
\end{abstract}


\end{opening}
\section{What is Compressible MHD Turbulence?}
It is well known that molecular clouds are magnetized 
with compressible magnetic
turbulence determining most of their properties (see
Elmegreen \& Falgarone 1996, Stutzki 2001).
Star formation (see McKee \& Tan 2002, Elmegreen 2002, 
Pudritz 2001), cloud chemistry (see Falgarone 1999 and
references therein), shattering and
coagulation of dust (see Lazarian \& Yan 2003 and references therein) 
are examples
of processes for which knowledge of turbulence is absolutely
essential.    
There are many excellent reviews that deal the theory
of turbulent molecular clouds and numerical simulations
(see Falgarone 1999,  
Vazquez-Semadeni et al. 2000, Mac Low \& Klessen 2003). 
This {\it short} review is focused on 
the recently uncovered basic properties of MHD turbulence 
and their implications for  
understanding of molecular clouds\footnote{
     It is not possible to cite all the important papers in the area
     of MHD turbulence and turbulent molecular clouds. 
     An incomplete list of the references in a 
     recent review on the statistics of
     MHD turbulence by Cho, Lazarian \& Vishniac (2003a; 
     henceforth CLV03a) includes about two hundred entries. The list
     of references in the molecular cloud review by Mac Low \& Klessen (2003)
     is even longer. }. We also briefly deal with
recovering of the 3D statistics of turbulent velocity from
observations, which is a theoretical problem in itself. A more
detailed discussion of MHD turbulence can be found in the
companion review by Chandran, while observational aspects of turbulence
in molecular clouds are discuss in the review by Falgarone (this volume).

 Why do we expect molecular clouds to be turbulent?
A fluid of viscosity $\nu$ becomes turbulent when the rate of viscous 
dissipation, which is  $\sim \nu/L^2$ at the energy injection scale $L$, 
is much smaller than
the energy transfer rate $\sim V_L/L$, where $V_L$ is the velocity dispersion
at the scale $L$. The ratio of the two rates is the Reynolds number 
$Re=V_LL/\nu$. In general, when $Re$ is larger than $10-100$
the system becomes turbulent. Chaotic structures develop gradually as 
$Re$ increases,
and those with $Re\sim10^3$ are appreciably less chaotic than those
with $Re\sim10^8$. Observed features such as star forming clouds are
very chaotic for $Re>10^8$. 
This makes it difficult to simulate realistic turbulence. 
The currently available
3D simulations containing 512 grid cells along each side
can have $Re$ up to $\sim O(10^3)$
and are limited by their grid sizes. 
Therefore, it is essential to find ``{\it scaling laws}" in order to
extrapolate numerical calculations ($Re \sim O(10^3)$) to
real astrophysical fluids ($Re>10^8$). 
We show below that even with its limited resolution, numerics is a great 
tool for {\it testing} scaling laws.

Kolmogorov theory provides a scaling law for {\it incompressible} 
{\it non}-magnetized hydrodynamic turbulence (Kolmogorov 1941).
This law provides a statistical relation
between the relative velocity $v_l$ of fluid elements and their separation
$l$, namely, $v_l\sim l^{1/3}$.  An equivalent description is to 
express spectrum $E(k)$
as a function of wave number $k$ ($\sim 1/l$).
The two descriptions are related by $kE(k) \sim v_l^2$. The famous
Kolmogorov spectrum is  $E(k)\sim k^{-5/3}$. The applications of 
Kolmogorov theory range from engineering research to
meteorology (see Monin \& Yaglom 1975) but its astrophysical
applications are poorly justified and the application
of the Kolmogorov theory can lead to erroneous conclusions
(see reviews by Lazarian et al.
2003 and Lazarian \& Yan 2003)

Let us consider {\it incompressible} MHD turbulence first.
There have long been understanding that the MHD turbulence
is anisotropic
(e.g. Shebalin et al.~1983). Substantial progress has been achieved
recently by Goldreich \& Sridhar (1995; hereafter GS95), who made an
ingenious prediction regarding relative motions parallel and
perpendicular to magnetic field {\bf B} for incompressible
MHD turbulence. 
An important observation that leads to understanding of the GS95
scaling is that magnetic field cannot prevent mixing motions
of magnetic field lines if the motions
are perpendicular to the magnetic field. Those motions will cause, however,
waves that will propagate along magnetic field lines.
If that is the case, 
the time scale of the wave-like motions along the field, 
i.e. $\sim l_{\|}/V_A$,
($l_{\|}$ is the characteristic size of the perturbation along 
the magnetic field and 
$V_A=B/\sqrt{4 \pi \rho}$ is 
the local Alfven speed) will be equal to the hydrodynamic time-scale, 
$l_{\perp}/v_l$, 
where $l_{\perp}$ is the characteristic size of the perturbation
perpendicular to the magnetic field.
The mixing motions are 
hydrodynamic-like\footnote{
        Recent simulations (Cho et al.~2003c) suggest that
        perpendicular mixing is indeed efficient for
        mean magnetic fields of up to the equipartition value.}.
They obey Kolmogorov scaling,
$v_l\propto l_{\perp}^{1/3}$, because incompressible turbulence is assumed. 
Combining the two relations above
we can get the GS95 anisotropy, $l_{\|}\propto l_{\perp}^{2/3}$ 
(or $k_{\|}\propto k_{\perp}^{2/3}$ in terms of wave-numbers).
If  we interpret $l_{\|}$ as the eddy size in the direction of the 
local 
magnetic field\footnote{Incidentally, if we identify the size
of molecular cloud with the longer dimension of
an eddy, we shall get the Larson relations $v_l\sim l^{1/2}$ (Larson 1981) for
velocity and cloud size.}
and $l_{\perp}$ as that in the perpendicular directions,
the relation implies that smaller eddies are more elongated.

GS95 predictions have been confirmed 
numerically (Cho \& Vishniac 2000; Maron \& Goldreich 2001;
Cho, Lazarian \& Vishniac 2002a, hereafter CLV02a; see also CLV03a); 
they are in good agreement with observed and inferred astrophysical spectra 
(see CLV03a). However, the
GS95 model considered incompressible MHD, but the media in
molecular clouds is {\it highly compressible}. Does any part
of GS95 model survives?
Literature on the properties of compressible MHD is very rich (see reviews
by Pouquet 1999, Cho \& Lazarian 2003 and references therein).
Higdon (1984) theoretically studied density fluctuations
in the interstellar MHD turbulence.
Matthaeus \& Brown (1988) studied nearly incompressible MHD at low Mach
number and Zank \& Matthaeus (1993) extended it. In an important paper
Matthaeus et al.~(1996) numerically
explored anisotropy of compressible MHD turbulence. However, those
papers do not provide universal scalings of the GS95 type.

The complexity of the
compressible magnetized turbulence with magnetic field made some
researchers believe that the phenomenon is too complex to expect any
universal scalings for molecular cloud research.
High coupling of compressible
and incompressible motions is often quoted to justify this 
point of view.

Below we shall provide arguments that are suggestive that the fundamentals
of compressible MHD can be understood and successfully applied to
molecular clouds.

In molecular clouds the regular magnetic field is comparable
with the fluctuating one. Therefore 
for most part of our discussion, we shall discuss results
obtained for 
$\delta V \sim \delta B/\sqrt{4 \pi \rho} \sim B_0/\sqrt{4 \pi \rho}$,
where $\delta B$ is the r.m.s. strength of the random magnetic field.

\section{Does the Decay of MHD Turbulence Depend on Compressibility? }
Turbulent support of molecular clouds (see review by McKee 1999) critically
depends on the rate of turbulence decay. 
For a long time magnetic fields were thought to be the means of
decreasing dissipation of turbulence. 
Numerical calculations by
Mac Low et al. (1998) and Stone et al. (1998) indicated that 
compressible MHD turbulence
decays as fast as the hydrodynamic turbulence. This gives rise to a 
belief that it is the compressibility that 
is responsible for the rapid decay of MHD turbulence.  

This point of view has been recently subjected to scrutiny in 
Cho \& Lazarian (2002, 2003a, henceforth CL02 and
CL03, respectively).  
In these papers a
 technique of separating different MHD modes was developed and used
(see Fig.~1). 
This allowed us to follow how the energy was redistributed between
these modes.
\begin{figure*}
  \includegraphics[width=0.90\textwidth]{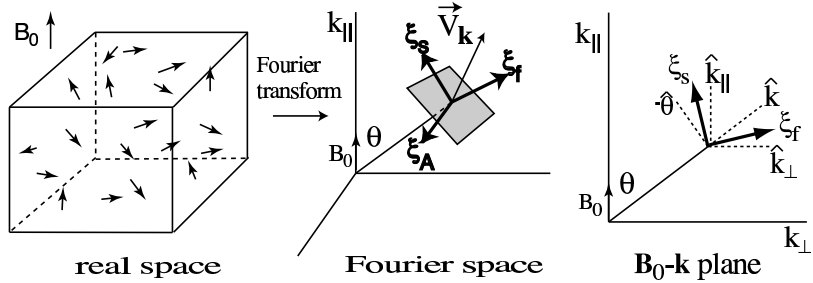}
  \caption{
      Separation method. We separate Alfven, slow, and fast modes in Fourier
      space by projecting the velocity Fourier component ${\bf v_k}$ onto
      bases ${\bf \xi}_A$, ${\bf \xi}_s$, and ${\bf \xi}_f$, respectively.
      Note that ${\bf \xi}_A = -\hat{\bf \varphi}$. 
      Slow basis ${\bf \xi}_s$ and fast basis ${\bf \xi}_f$ lie in the
      plane defined by ${\bf B}_0$ and ${\bf k}$.
      Slow basis ${\bf \xi}_s$ lies between $-\hat{\bf \theta}$ and 
      $\hat{\bf k}_{\|}$.
      Fast basis ${\bf \xi}_f$ lies between $\hat{\bf k}$ and 
      $\hat{\bf k}_{\perp}$. 
}
\label{fig_separation}
\end{figure*}

How is this idealized incompressible model related to the actual
turbulence in molecular clouds? 
Compressible MHD turbulence is a highly non-linear phenomenon
and it has been thought that 
different types of perturbations or modes (Alfven, slow and fast)
in compressible media are strongly coupled. Nevertheless,
one may question whether this is true.
A remarkable feature of the GS95 model is that
Alfven perturbations cascade to small scales over just one wave
period, while the other non-linear interactions require more time.
Therefore one might expect
that the non-linear interactions with other types of waves
should affect Alfvenic cascade only marginally. 
Moreover, since the Alfven waves are incompressible, the properties
of the corresponding cascade may not depend on the sonic Mach number.

The generation of compressible motions 
(i.e. {\it radial} components in Fourier space) 
{}from Alfvenic turbulence
is a measure of mode coupling.
How much energy in compressible motions is drained from Alfvenic cascade?
According to closure calculations (Bertoglio, 
Bataille, \& Marion 2001; see also Zank \& Matthaeus 1993),
the energy in compressible modes in {\it hydrodynamic} turbulence scales
as $\sim M_s^2$ if $M_s<1$.
CL03 conjectured that this relation can be extended to MHD turbulence
if, instead of $M_s^2$, we use
$\sim (\delta V)_{A}^2/(a^2+V_A^2)$. 
(Hereinafter, we define $V_A\equiv B_0/\sqrt{4\pi\rho}$, where
$B_0$ is the mean magnetic field strength.) 
However, since the Alfven modes 
are anisotropic, 
this formula may require an additional factor.
The compressible modes are generated inside the so-called
Goldreich-Sridhar cone, which takes up $\sim (\delta V)_A/ V_A$ of
the wave vector space. The ratio of compressible to Alfvenic energy 
inside this cone is the ratio given above. 
If the generated fast modes become
isotropic (see below), the diffusion or, ``isotropization'' of the
fast wave energy in the wave vector space increase their energy by
a factor of $\sim V_A/(\delta V)_A$. This  results in
\begin{equation}
  \frac{ (\delta V)_{rad}^2 }{ (\delta V)_A^2 }   \sim
 \left[ \frac{ V_A^2 + a^2 }{ (\delta V)^2_A } 
        \frac{ (\delta V)_A }{ V_A }   \right]^{-1},
\label{eq_high2}
\end{equation}
where $(\delta V)_{rad}^2$ and $(\delta V)_{A}^2$ are energy
of compressible  and Alfven modes, respectively.
Eq.~(\ref{eq_high2}) suggests that the drain of energy from
Alfvenic cascade is marginal\footnote{
	The marginal generation of compressible 
        modes is in agreement with 
        earlier studies by Boldyrev et al. (2002b) and 
        Porter, Pouquet, \& Woodward (2002),
        where the
        velocity was decomposed into a potential component
        and a solenoidal component. A recent study by
	Vestuto, Ostriker \& Stone 
(2003) is also consistent with this conclusion. }
when the amplitudes of perturbations
are weak, i.e. $(\delta V)_A \ll  V_A$. Results of calculations
shown in Fig.~2 support the theoretical predictions.

We may summarize this issue in the following way. For the incompressible
motions to decay fast, there is no requirement of coupling with 
compressible motions\footnote{
   The reported (see Mac Low et al.~1998) decay of the {\it total}
   energy of turbulent motions $E_{tot}$ follows $t^{-1}$ which can
   be understood if we account for the fact that the energy is being
   injected at the scale smaller than the scale of the system. Therefore
   some energy originally diffuses to larger scales through the inverse 
   cascade. Our calculations (Cho \& Lazarian, unpublished), 
   stimulated by illuminating 
   discussions with Chris McKee, show that if this energy 
   transfer is artificially
   prevented by injecting the energy on the scale of the computational box, 
   the scaling of $E_{tot}$  becomes closer to $t^{-2}$.}. 
The marginal coupling of the compressible 
and incompressible modes allows to study modes
separately. 
 
\begin{figure*}
  \includegraphics[width=0.34\textwidth]{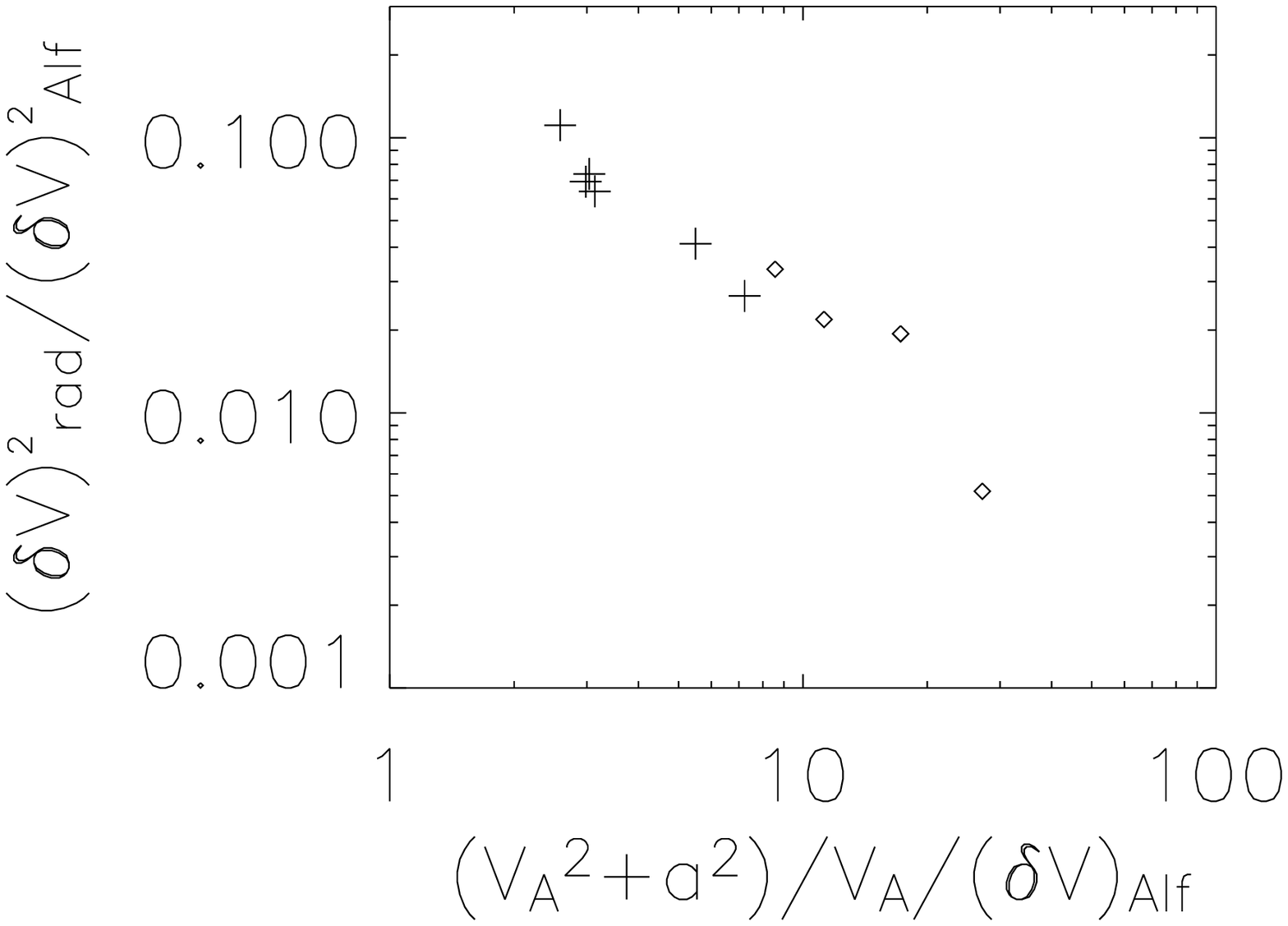}
\hfill
  \includegraphics[width=0.24\textwidth]{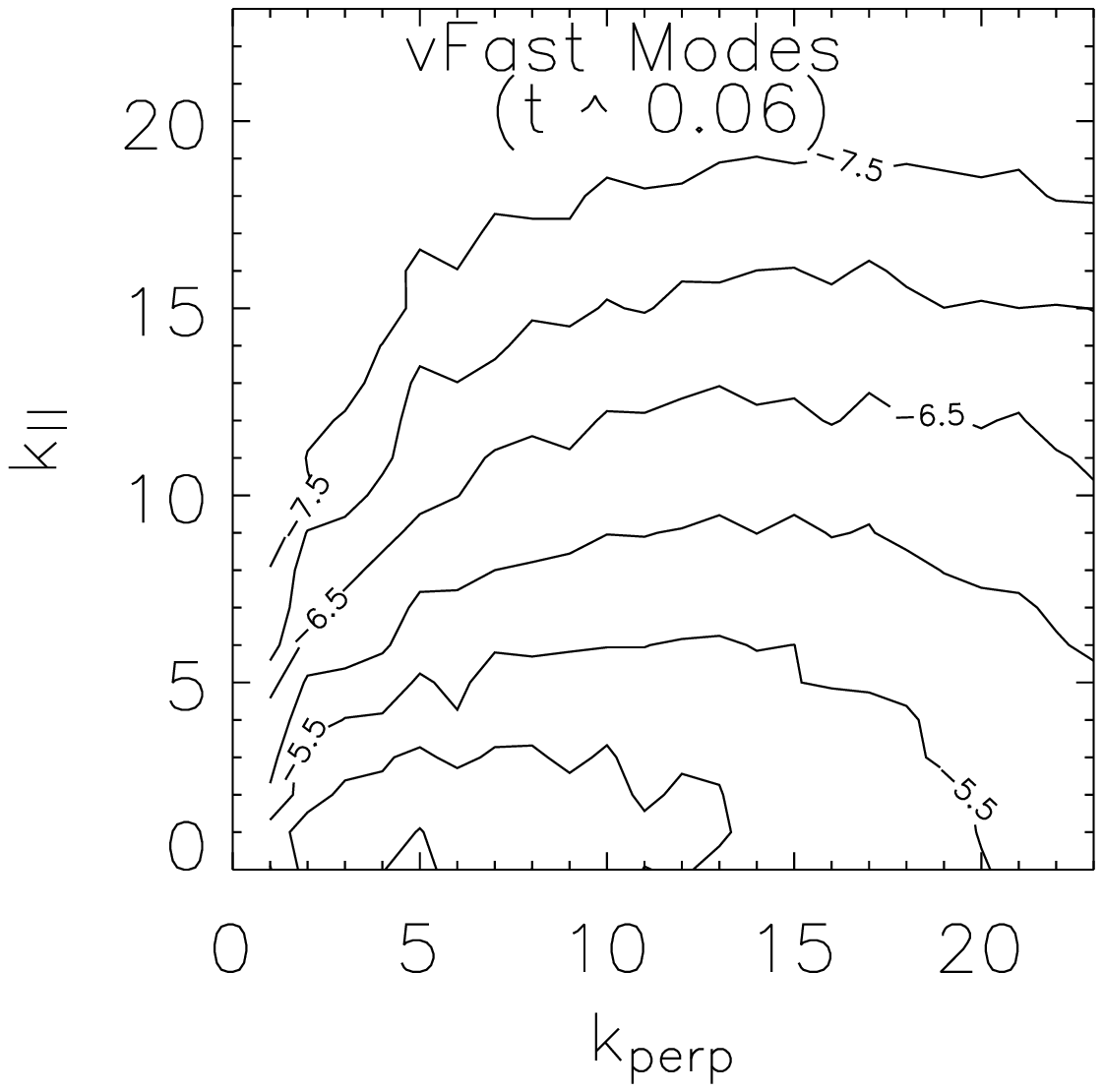}
\hfill
  \includegraphics[width=0.3\textwidth]{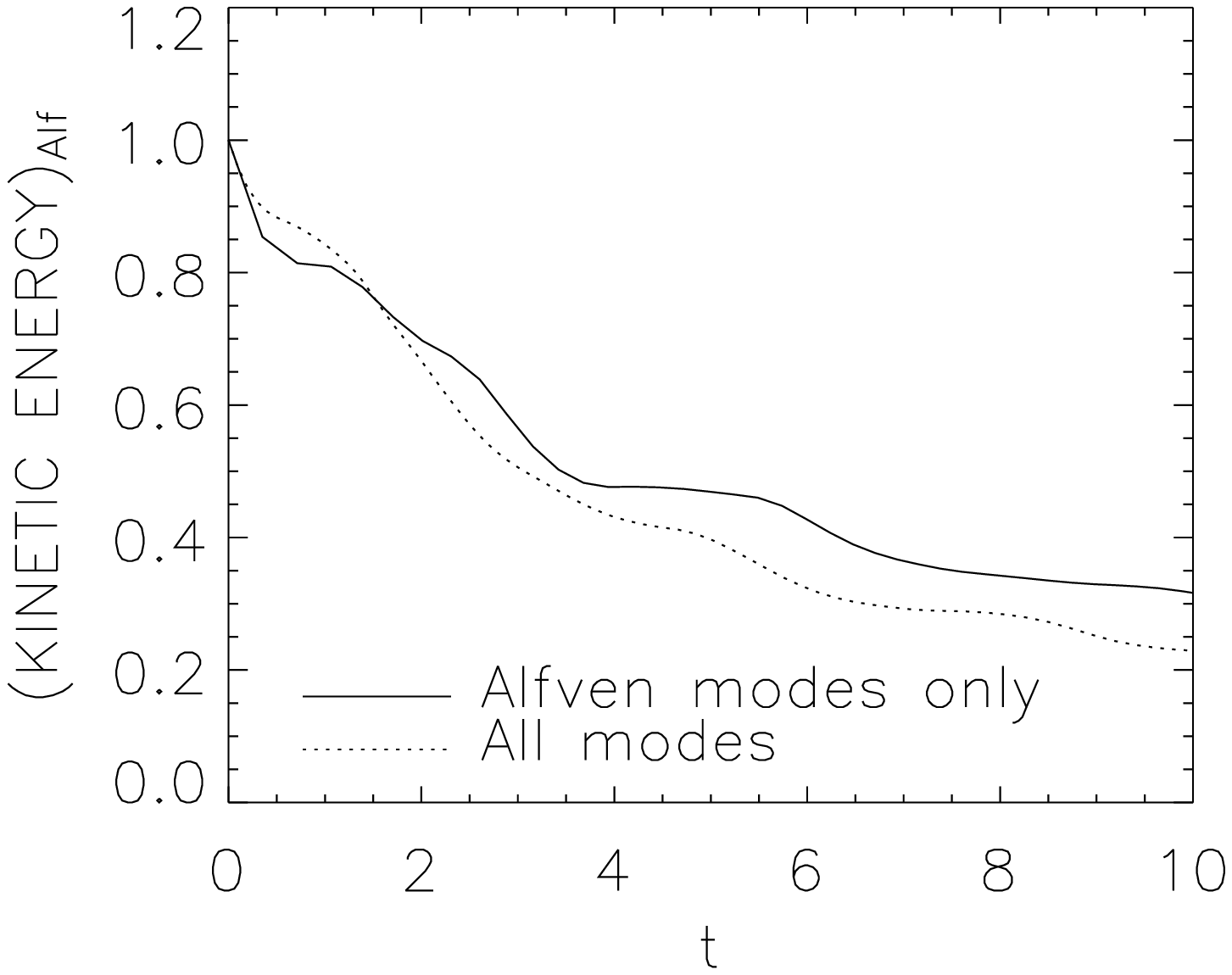}
  \caption{
      Mode coupling studies.
    (a){\it left:}  Square of the r.m.s. velocity of the compressible modes.
        We use $144^3$ grid points. Only Alfven modes are allowed
        as the initial condition.
        ``Pluses'' are for low $\beta$ cases ($0.02 \leq \beta \leq 0.4$).
        ``Diamonds'' are for high  $\beta$ cases ($1 \leq \beta \leq 20$).
    (b){\it middle:} Generation of fast modes. Snapshot is taken at t=0.06 from
        a simulation (with $144^3$ grid points) 
        that started off with Alfven modes only.
        Initially, $\beta$ (ratio of gas to magnetic pressure, $P_g/P_{mag}$) 
          $=0.2$ and 
          $M_s$ (sonic Mach number) $\sim 1.6$.
    (c){\it right:} Comparison of decay rates.
        Decay of Alfven modes is not much affected by other 
       (slow and fast) modes. We use $216^3$ grid points.
        Initially, $\beta=0.02$ and 
        $M_s\sim 4.5$ for the solid line and 
        $M_s\sim 7$ for the dotted line. 
        Note that initial data are, in some sense, identical for
        the solid and the dotted lines.
        The sonic Mach number for the solid line is smaller
        because we removed fast and slow modes from the initial data before
        the decay simulation.
        For the dotted line, we did {\it not} remove any modes from the
        initial data. From CL03.
}
\label{fig_coupling}
\end{figure*}

\section{Is Turbulence in Molecular Clouds Really Messy?}
Is it feasible to obtain scaling relations for the compressible MHD
turbulence?
Some hints about effects of compressibility can be inferred from 
the GS95 seminal paper. More discussion was
presented in Lithwick \& Goldreich (2001), which deals with electron
density fluctuations in the gas pressure dominated plasma, 
i.e.  in high $\beta$ regime ($\beta\equiv P_{gas}/P_{mag}\gg 1$). 
The incompressible regime 
corresponds to $\beta\rightarrow \infty$, so it is natural
to expect that for $\beta\gg 1$ the GS95 picture would
persist. Lithwick \&
Goldreich (2001) also speculated that for low $\beta$ plasmas the GS95
scaling of slow modes may be applicable. 
A detailed 
study of compressible mode scalings  is given in CL02 and CL03. 

Our considerations above about the mode coupling can guide us
to predict the mode scaling. Indeed,
if Alfven cascade evolves on its own, it is natural to assume that 
slow modes exhibit the GS95 scaling.
Indeed, slow modes in gas 
pressure dominated environment (high $\beta$ plasmas) are
similar to the pseudo-Alfven modes in incompressible regime 
(see GS95; Lithwick \& Goldreich 2001). The latter modes do follow
the GS95 scaling. 
In magnetic pressure dominated environments (low $\beta$ plasmas), 
slow modes are density perturbations propagating with the
sound speed $a$ parallel to the mean magnetic field. 
Those perturbations are essentially
static for $a\ll V_A$. 
Therefore Alfvenic turbulence is expected to mix density
perturbations as if they were passive scalar. This also induces the
GS95 spectrum.

The fast waves in low $\beta$ regime propagate at $V_A$ irrespectively
of the magnetic field direction. 
In high $\beta$ regime, the properties of fast modes are similar, 
but the propagation speed is the sound speed $a$.
Thus the mixing motions induced by Alfven waves should marginally
affect the fast wave
cascade. It is expected to
be analogous to the acoustic wave cascade and hence be isotropic.

Results of numerical calculations from Cho \& Lazarian (CL03) 
for magnetically dominated media similar to that in molecular 
clouds are
shown in Fig.~3. They support theoretical considerations above. 

\begin{figure*}
  \includegraphics[width=0.90\textwidth]{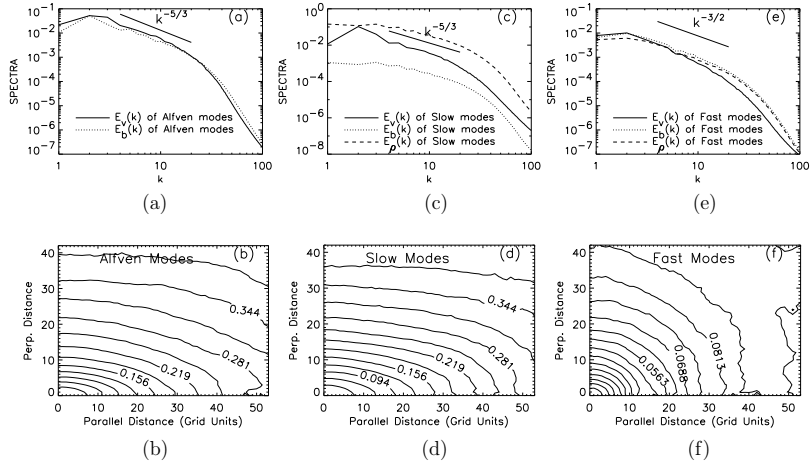}
  \caption{          $M_s\sim 2.2$, $M_A\sim 0.7$, $\beta\sim 0.2$,
           and $216^3$ grid points.
          (a) Spectra of Alfv\'en modes follow a Kolmogorov-like power
              law.
          (b) Eddy shapes
              (contours of same second-order structure function, $SF_2$)
              for velocity of Alfv\'en modes
              shows anisotropy similar to the GS95
         ($r_{\|}\propto r_{\perp}^{2/3}$ or $k_{\|}\propto
                                              k_{\perp}^{2/3}$).
              The structure functions are measured in directions
              perpendicular or
              parallel to the local mean magnetic field in real space.
              We obtain real-space velocity and magnetic fields
              by inverse Fourier transform of
              the projected fields.
          (c) Spectra of slow modes also follow a Kolmogorov-like power
              law.
          (d) Slow mode velocity shows anisotropy similar to the GS95.
              We obtain contours of equal $SF_2$ directly in real space
              without going through the projection method,
              assuming slow mode velocity is nearly parallel to local
              mean magnetic field in low $\beta$ plasmas.
          (e) Spectra of fast modes are compatible with
              the IK spectrum.
          (f) The magnetic $SF_2$ of
              fast modes shows isotropy.  From CL02
    } 
\label{fig_M2}
\end{figure*}

Why would we care about those scalings? How wrong is it to use Kolmogorov
scalings instead? Dynamics, chemistry and physics of molecular clouds
(see Falgarone 1999) 
presents a complex of problems for which the exact scalings may be
required to a different degree. 
If we talk about dynamics of interstellar dust or
propagation of
cosmic rays, one {\it must} account for the actual scalings and couplings
of different modes (see reviews by 
Lazarian \& Yan 2003, Lazarian et al. 2003). There are other problems,
e.g. turbulent heat transport where the exact scaling of modes seems
to be less important (Cho et al. 2003). 

\section{What is the effect of partial ionization?}

Our considerations above assume that the gas is fully ionized. Molecular 
clouds are partially ionized. What is the consequence of this?
An obvious effect of neutrals 
is that they do not follow magnetic field lines and thus produce viscosity.

In hydrodynamic turbulence viscosity sets a minimal scale for
motion, with an exponential suppression of motion on smaller
scales.  Below the viscous cutoff the kinetic energy contained in a 
wavenumber band is 
dissipated at that scale, instead of being transferred to smaller scales.
This means the end of the hydrodynamic cascade, but in MHD turbulence
this is not the end of magnetic structure evolution.  For 
viscosity much larger than resistivity, there will be a broad range of
scales where viscosity is important but resistivity is not.  
On these
scales magnetic field structures will be created 
by the shear from non-damped turbulent motions, which
amounts essentially to the shear from the smallest undamped scales.
Indeed, this new regime of turbulence has been discovered
(see Fig.~4)!

\begin{figure*}
  \includegraphics[width=0.49\textwidth]{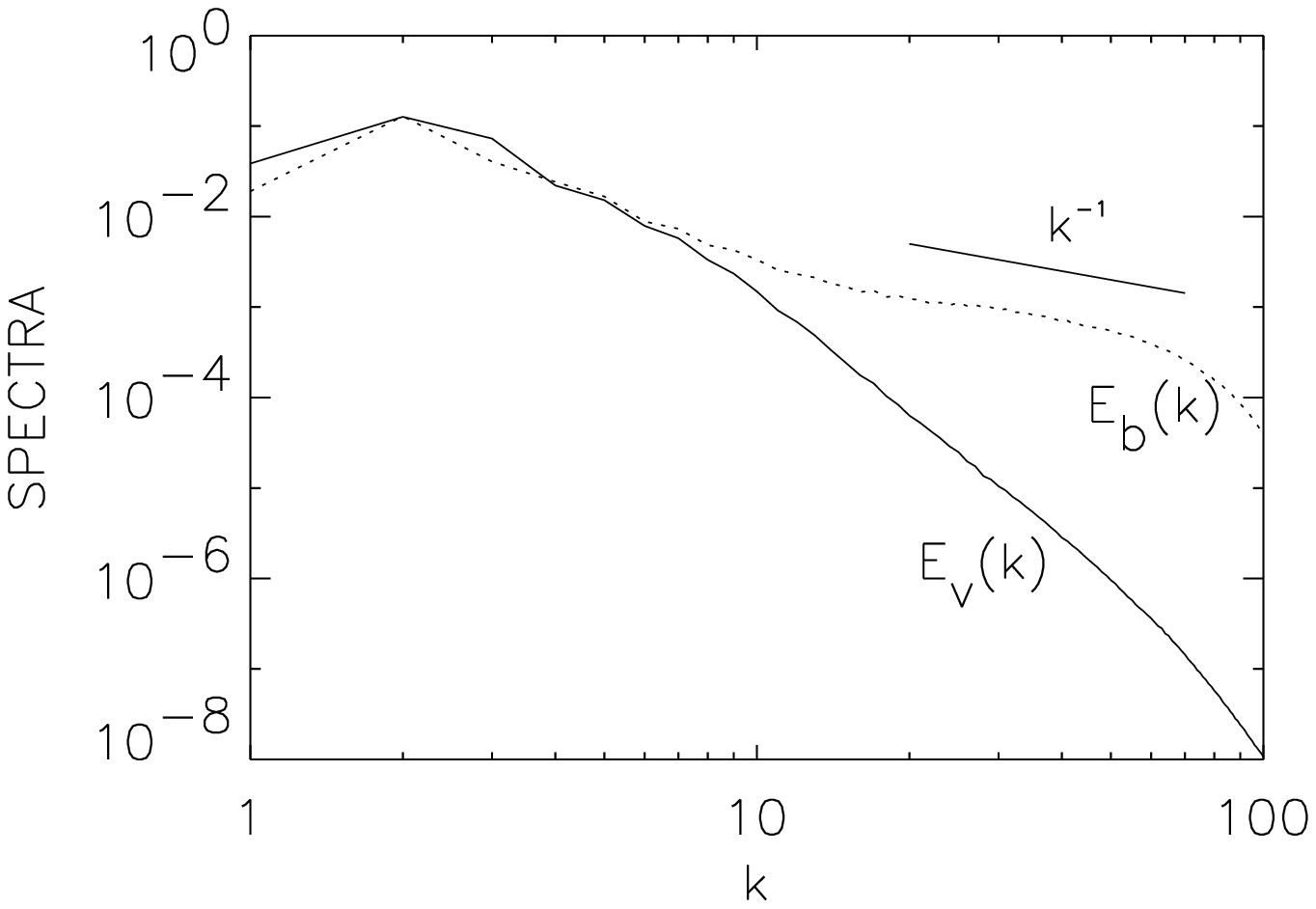}
\hfill
  \includegraphics[width=0.49\textwidth]{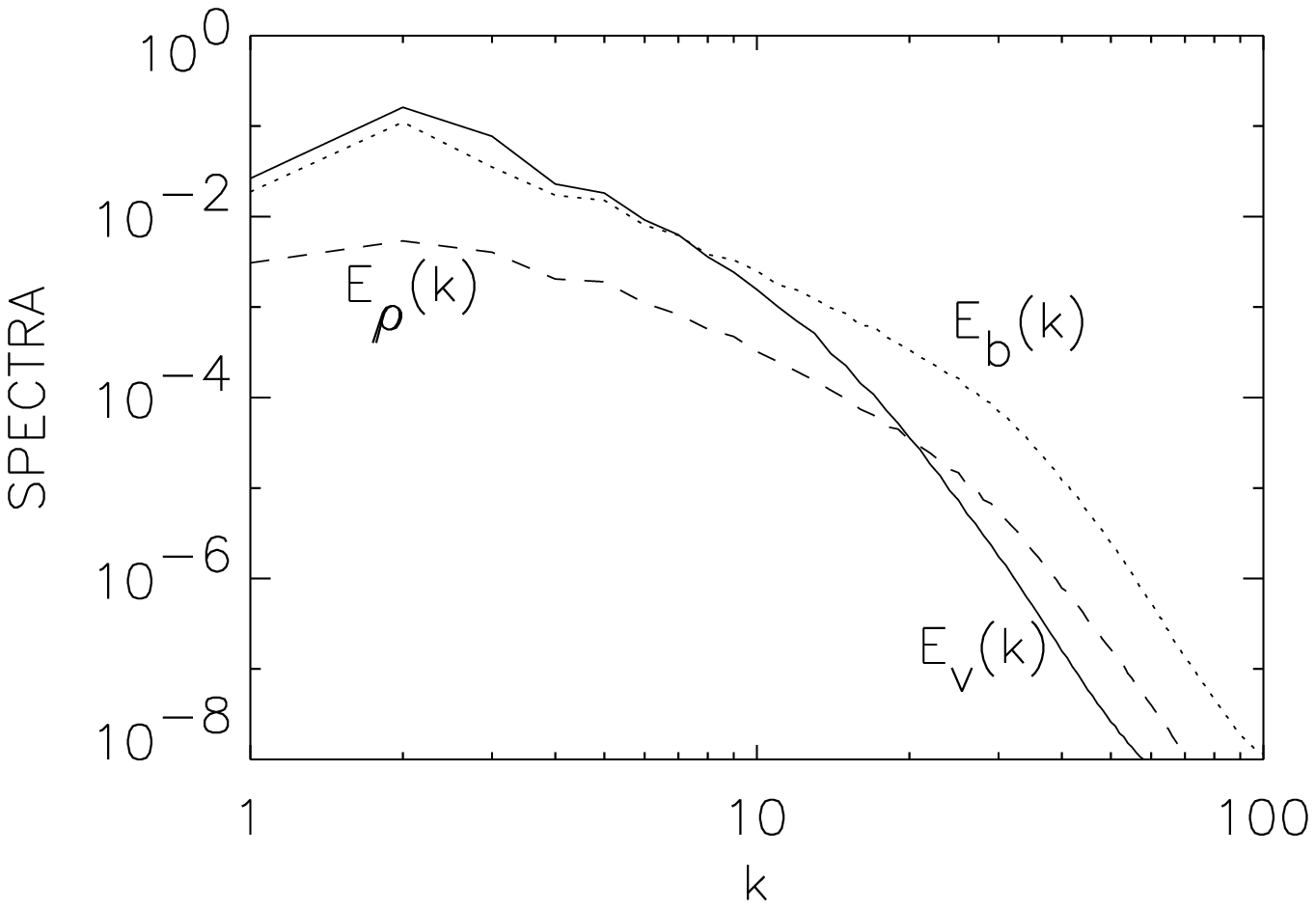}
  \caption{
      Viscous damped regime (viscosity $>$ magnetic diffusivity).
      Due to large viscosity, velocity damps after $k\sim10$.
    (a) {\it Left:} Incompressible case with $384^3$ grid points. 
        Magnetic spectra show a shallower slope ($E_b(k)\propto k^{-1}$)
        below the velocity damping scale.
        We achieve a very small magnetic diffusivity through the use
        hyper-diffusion.
       {}From Cho, Lazarian, \& Vishniac (2002b).
    (b) {\it Right:} Compressible case with $216^3$ grid points.
        Magnetic and density spectra show structures below the
        velocity damping scale at $k\sim10$.
        The structures are less obvious than the incompressible case
        because it is relatively hard to
        achieve very small magnetic diffusivity in the compressible run.
        From CL03.}
\label{fig_viscous}
\end{figure*}

A theoretical model for this new regime  
can be found in Lazarian, Vishniac, \& Cho (2003; hereafter LVC03). 
It explains the spectrum $E(k)\sim k^{-1}$ as a cascade of magnetic
energy to small scales under the influence of shear at the 
marginally damped scales. 
     The spectrum is similar to that of the viscous-convective range of
     passive scalar in hydrodynamic turbulence (see, for example,
     Batchelor 1959; Lesieur 1990), although the study in Lazarian,
     Vishniac, \& Cho (LVC03) suggests that the physical origin of it 
     is different.
     A study confirming that the $k^{-1}$ spectrum
        is not a bottleneck effect is presented in 
        Cho, Lazarian, \& Vishniac (2003b).
The mechanism is based on the solenoidal
motions and therefore the compressibility should not alter the physics
of this regime of turbulence.  

For compressible simulations (see Fig. 4) the inertial 
range is much smaller due to numerical reasons, but it is clear that
the viscosity-damped regime of MHD turbulence persists. 
The magnetic fluctuations,
however, compress the gas and thus cause fluctuations in density.
These density fluctuations may have important
and yet unexplored 
consequences for the small scale structure of the molecular clouds.
There are indications that small-scale structure is indeed present
in molecular gas (Marscher, Moore, \& Bania 1993) as well
as in other partially ionized phases of the ISM (see Heiles 1997).
Those structures can be produced by turbulence with a spectrum
substantially more shallow than the Kolmogorov one, e.g. with the
spectrum $E(k)\sim k^{-1}$ (see Deshpande 2000). 
Below the viscous scale the fluctuations of 
magnetic field obey the damped regime shown in Fig.~\ref{fig_viscous}b 
and produce 
density fluctuations. For typical Cold Neutral Medium gas (see
Draine \& Lazarian 1998), these
fluctuations can be as large as
 $\sim70$ AU and less for denser gas.

An important prediction in LVC03 is that the intermittency 
of magnetic structures increases with the decrease of the
scale. This prediction was confirmed by numerical simulations
in Cho, Lazarian \& Vishniac (2003), which showed that the
filling factor of magnetic field was decreasing with the
increase of the wavenumber. 

Does the effect of neutrals amounts only to emergence of the 
viscosity-damped regime of MHD turbulence? The answer to this
question is negative. Partial ionization provides the whole range
of new interesting effects. Some of them are studied in LVC03. 
Here we mention those.

First of all, it is clear that whether ions and neutrals act as
one fluid in molecular clouds depends on whether the eddy turnover
rate $t_{eddy}^{-1}\sim k v_k$ is longer or shorter than the rate
$t_{ni}^{-1}$ of neutral-ion collisions. If $t_{eddy}^{-1}>
t_{ni}^{-1}$, neutrals decouple from ions and
develop {\it hydrodynamic} Kolmogorov-type cascade. Indeed, the damping
rate for those hydrodynamic motions $t_{ni}^{-1}$ and below the
decoupling scale  
the hydrodynamic motions evolve without much hindrance from
magnetic field. Magnetic fields with entrained ions develop 
the viscosity-damped MHD cascade until ion-neutral collisional
rate gets longer than the dynamical rate of the intermittent
magnetic structures. After that the turbulence reverts to its
normal MHD cascade which involves only ions. 

If $t_{eddy}^{-1}<t_{ni}^{-1}$ up to the scale at which neutral
viscosity damps turbulent motions, the viscosity-damped regime
emerges at the scale where kinetic energy associated with
turbulent eddies is dissipated. Similarly to the earlier case
when the when ion-neutral collisions get insufficient to preserve
pressure confinement of the small scale magnetic filaments, outbursts of
ordinary ionic MHD turbulence will take place. The turbulence will
be intermittent both in time and space 
because of the disparity of time scales at which
turbulence evolves in the viscosity-damped and free ionic MHD
regimes. We plan to test those predictions with a two fluid MHD code.

\section{How Can One Study Turbulence in Molecular Clouds?}

Substantial advances in understanding of MHD turbulence in
ordinary and viscosity-damped regime make it most important to 
test the correspondence of theoretical expectations and observational
reality (see review by Ostriker 2003). 
Important advances in obtaining high resolution spectral 
data (see Falgarone et al. 2000) make the testing feasible. 

The turbulence spectrum characterizes the distribution of energy over spatial scales
and reflects the processes of dissipation and energy injection.
The GS95 model
would also gives the Kolmogorov spectrum of Alfvenic motions
if averaging over ${\bf k}$ directions
is performed. This stems from the fact that for the inertial range
$k_{\bot}\gg k_{\|}$ and therefore $k\sim
k_{\bot}$. 

It is known that studies of stochastic
density provide only indirect insight into turbulence. 
A stationary (not evolving) density distribution is indistinguishable
from the result of active turbulence. However,
Doppler shifted spectral lines carry information about turbulent 
velocity. The problem is that both velocity and density 
fluctuations contribute to the observed fluctuations. 
Therefore numerous attempts to study turbulence in diffuse interstellar
medium and molecular clouds using emission lines
(see Munch 1958, O'Dell 1986, O'Dell \& Castaneda 1987,
Miesch \& Bally 1994, reviews by Scalo 1987, Falgarone 1999) 
were facing uncertainty of what quantity was actually measured. 

Studies of the velocity field have been attempted at different
times with velocity centroids
(e.g. Munch 1958, Miesch, Scalo, \& Bally 1999,  
Ossenkopf \& Mac Low 2002, Miville-Deschenes et al. 2003). 
However, it has long been realized that the centroids are
affected, in general, by both velocity and density fluctuations
(see Stenholm 1989). A criterion of when centroids indeed
reflect the velocity statistics was obtained in Lazarian
\& Esquivel (2003, henceforth LE03). 
Without this
testing, it is not clear {\it a priori} what is actually
being measured\footnote{A more optimistic claim about utility of
centroids was obtained in Miville-Deschenes, Levrier \& 
Falgarone (2003) on the basis of experiments with Brownian
noise. However, to make the density in their experiments positively
defined the authors added additional large mean density. According
to the criterion in LE03 this made their centroids bound to be
dominated by velocity.}.

An important recent development is a quantitative description of
spectral data from turbulent media obtained in Lazarian
\& Pogosyan (2000, 2003 henceforth LP00 and LP03, respectively),
where the fluctuations
of intensity in channel maps were related to the statistics of velocity 
and density.  
LP00 introduced a new technique that was termed 
Velocity Channel Analysis (henceforth VCA). Within VCA 
the separation of velocity and density contributions is obtained
by changing the thickness of the analyzed slice of the Position-Position-
Velocity (PPV) data cube. 
The VCA was successfully tested numerically in Lazarian et al. (2001)
and Esquivel et al. (2003) and applied to the Small Magellanic Cloud
(Stanimirovic \& Lazarian 2002).

\begin{figure*}
{\centering \leavevmode 
 \includegraphics[width=0.35\textwidth]{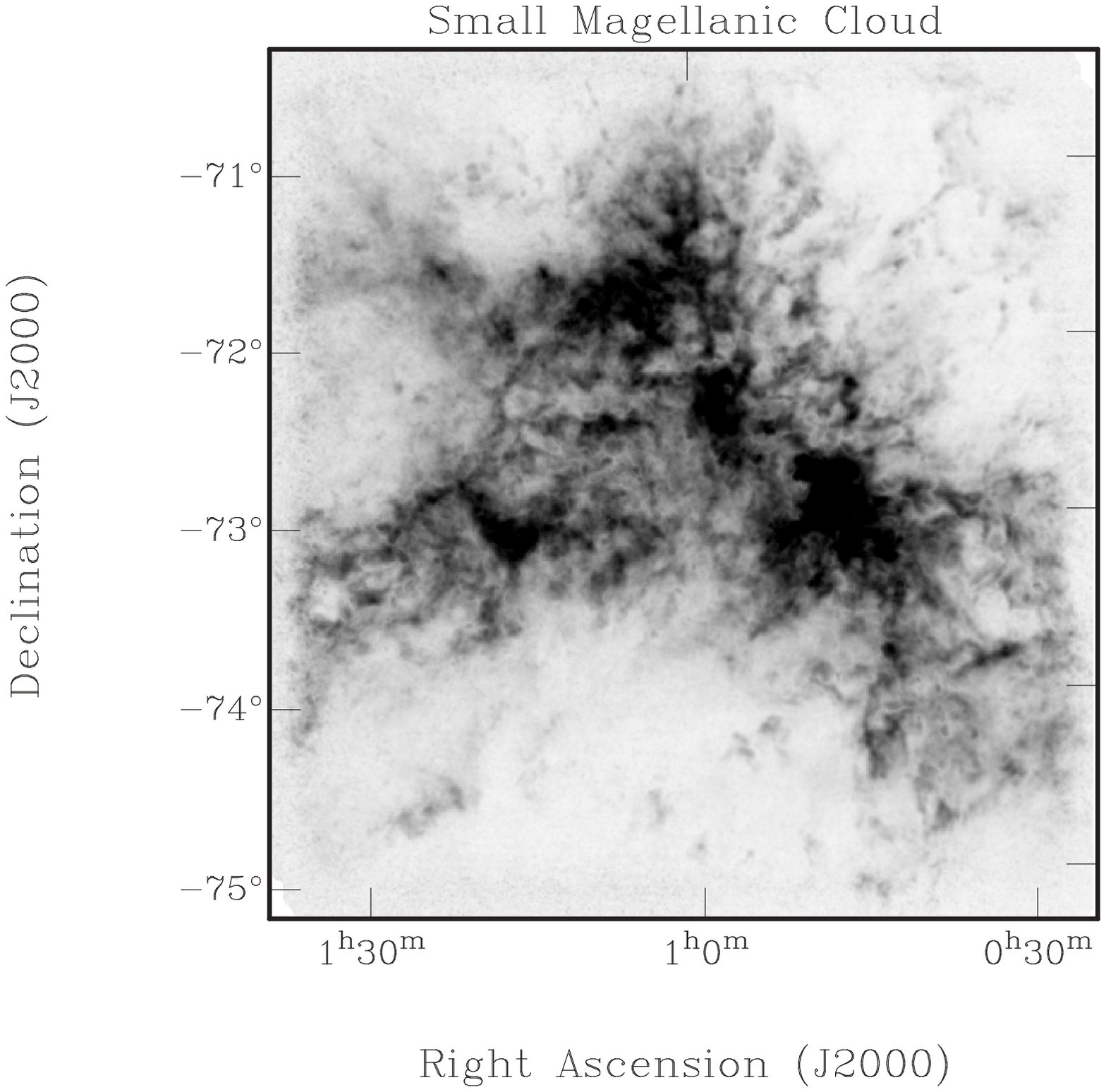} 
\hfil 
\ \includegraphics[width=0.50\textwidth]{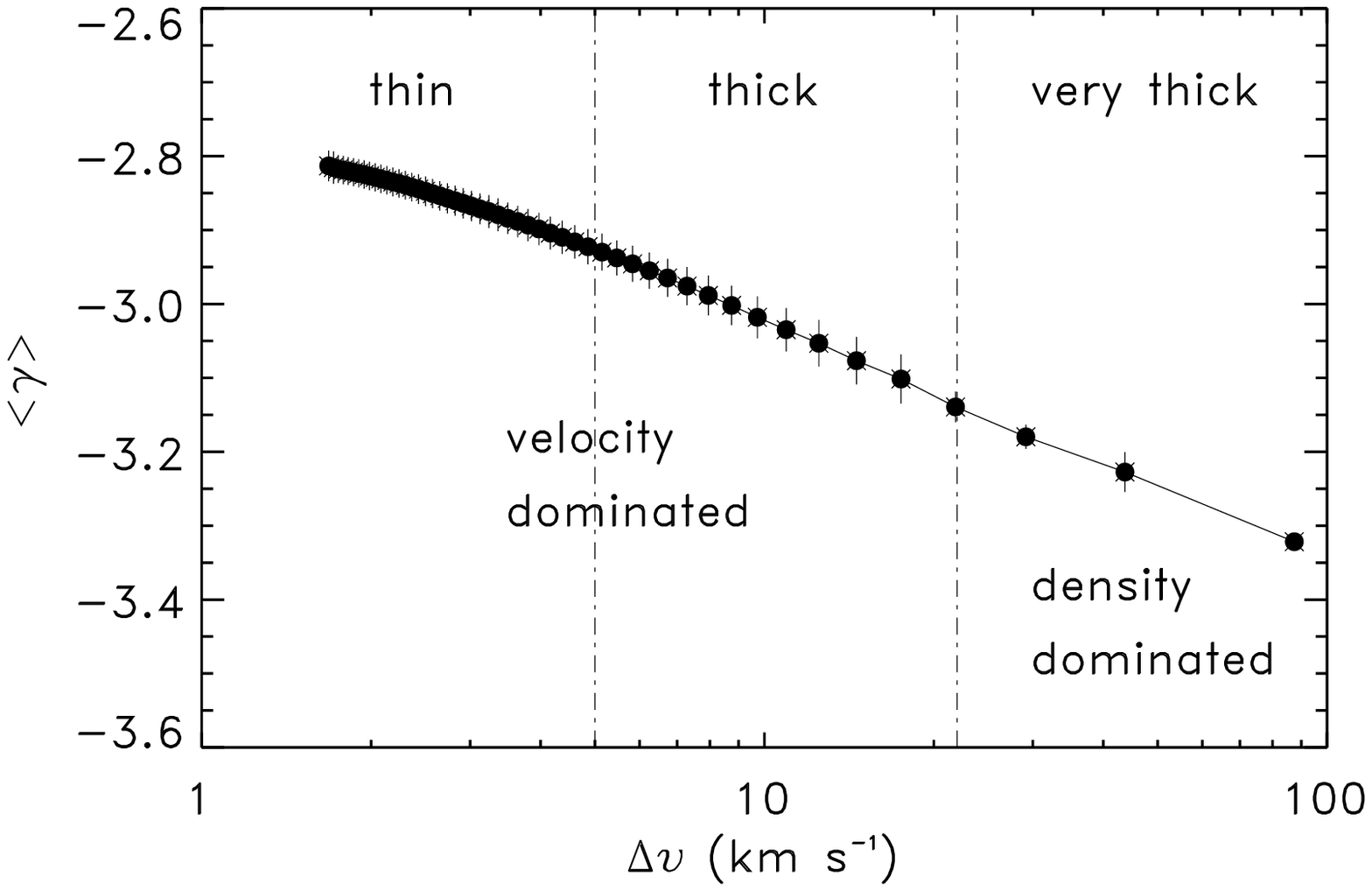}
}
{\centering \leavevmode
  \includegraphics[width=0.50\textwidth]{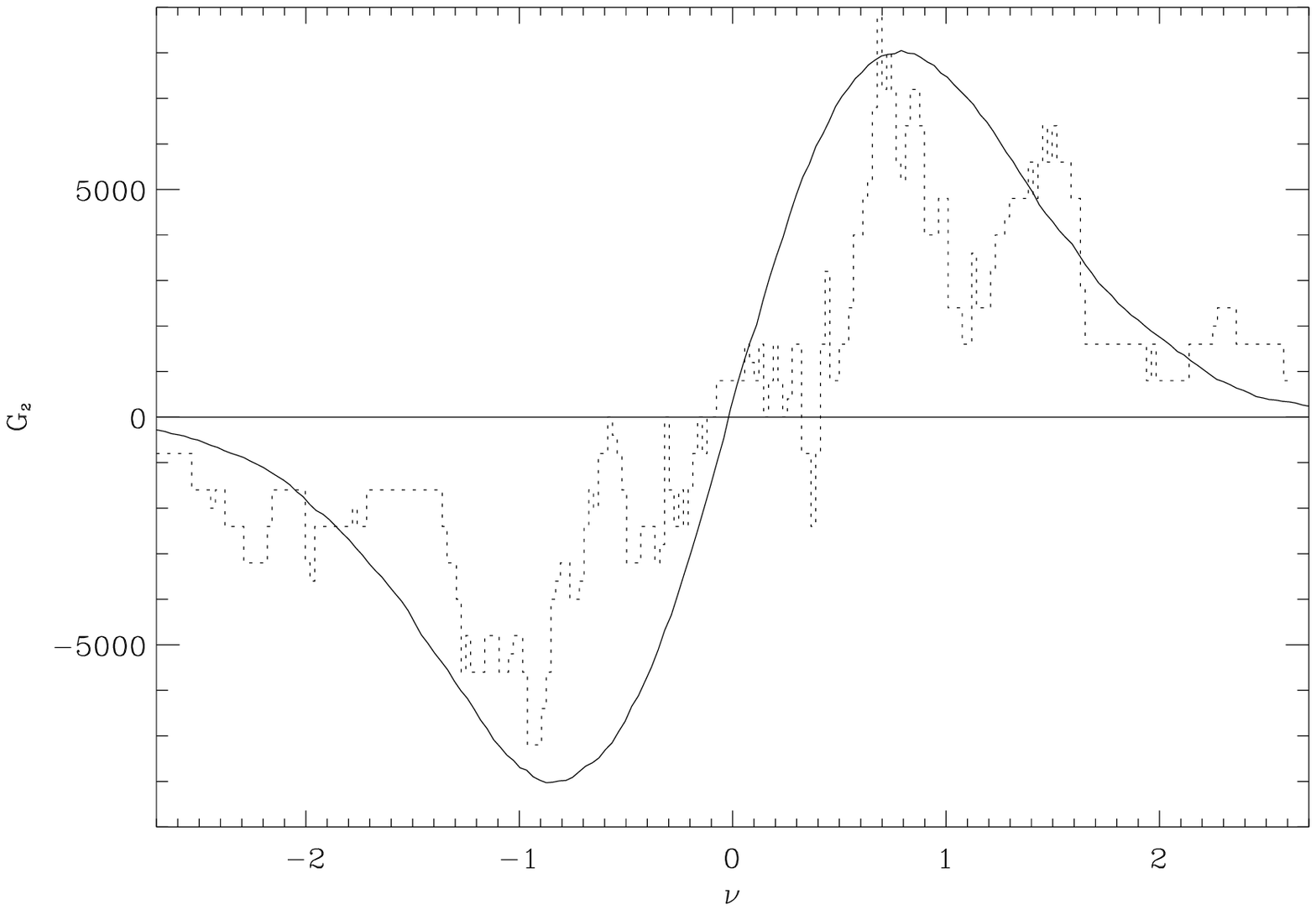} 
\hfil 
 \includegraphics[width=0.50\textwidth]{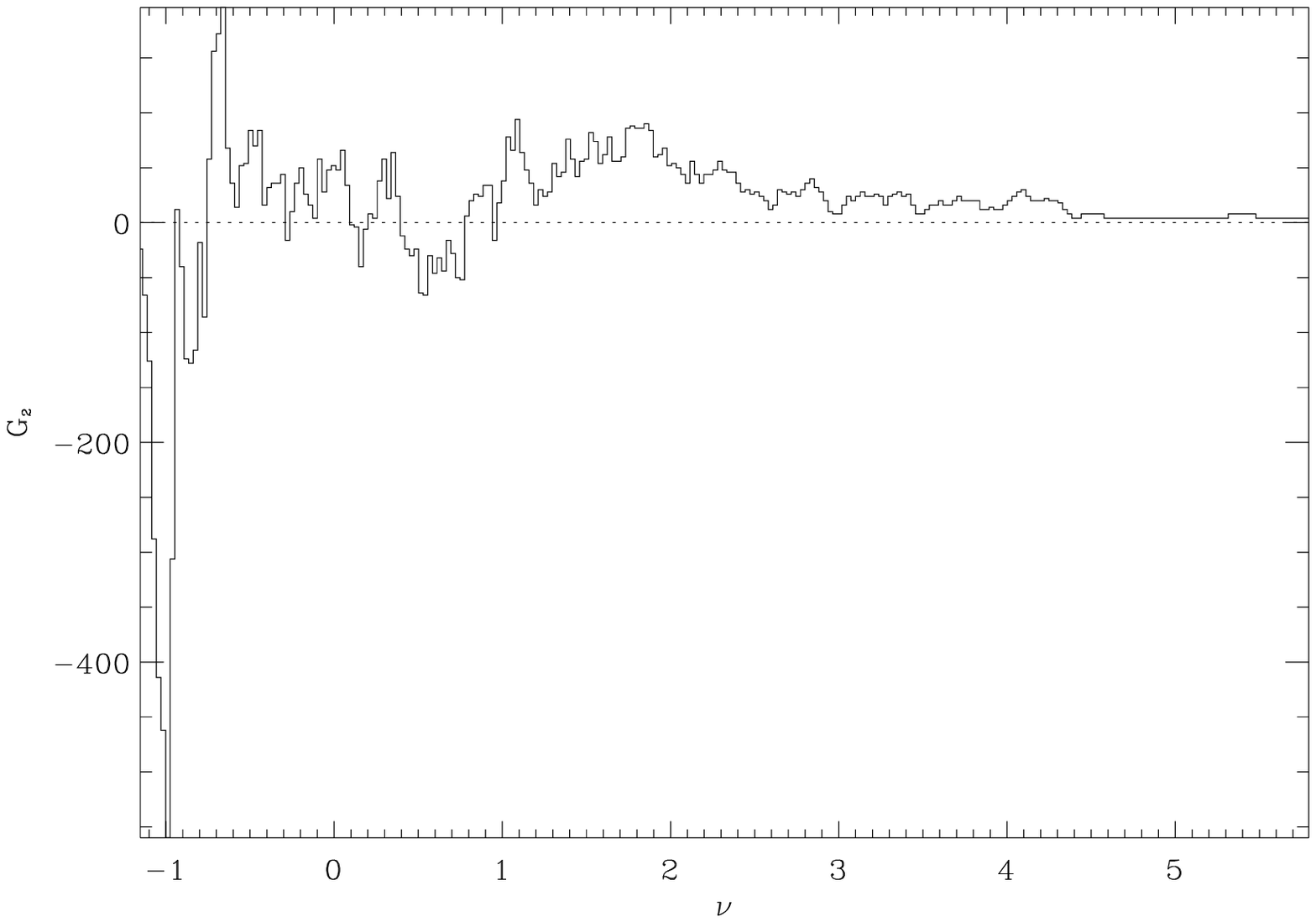}
} 
\caption{
 {\it Dealing with Observations: Testing and Applying New Statistical
 Tools.}  {\it Upper Left:} the 21~cm image of SMC, exhibiting strong
 density structure. According to our analysis this density
 has a spectrum close to that expected from MHD turbulence.  {\it
 Upper Right:} variations of 2-D 21~cm spectral
 slope with the velocity slice thickness (from Stanimirovic \&
 Lazarian 2001). The LP00 study predicts that the thick slice reflects
 the density statistics, while the thin slice is influenced by the
 velocity.  {\it Lower Left:} the 2D genus (which counts the number
 of holes versus the number of islands as the threshold contrast changes) 
 of the Gaussian distribution (smooth curve) against
 the genus for the isothermal compressible MHD simulations with Mach
 number 2.5 (dotted curve). Interstellar turbulence shows much more
 intermittency, with consequences for interstellar physics
 and chemistry.  {\it Lower Right:} the genus of HI
 distribution in SMC {\it that has the same power spectrum as the
dotted curve in the left panel} (Lazarian, Pogosyan
 \& Esquivel 2002). The genus (topology) of the distributions are very
 different!  }
\end{figure*}

LP03 accounts for the absorption
in the turbulent media. Thus it substantially extends the range of 
spectral lines that can be used for turbulence studies. Note that 
the importance of LP00 and LP03 studies is not limited to the
development of particular toolkit of how to interpret channel maps.
These works provide theoretical description that can be used within
different techniques. For instance, LP03 establishes limitations
on turbulence spectra that can be recovered by another promising
technique, namely, Spectral
Correlation Function (see Rosolowsky et al. 1999). 

Other important tools have also been studied recently.
Principal Component Analysis (PCA) 
of the emission data (Heyer \& Schloerb 1997,
Brunt \& Heyer 2002ab) provide a new promissing way to characterise
observations. Recent testing showed
that it provides statistics different from power spectra (Brunt et al. 2003). 
Wavelet analysis (see Gill \& Henriksen 1990), 
spectral correlation functions (see Rosolowsky et al. 1999,
Padoan, Rosolowsky \& Goodman 2001),
genus analysis (see Lazarian, Pogosyan \& Esquivel 2002) are other
statistical tools. They can provide statistics and insight 
complementary to power spectra.   
Synergy of different techniques should provide the necessary
insight into turbulence and enable the comparison of observational statistics
with theoretical expectations.

\section{What is the future of the field?}

There are whole classes of processes for which we are not sure even about
the sign of the effect, for instance, whether turbulence supports or
compresses molecular clouds. Or consider imbalanced turbulence,
i.e. the turbulence where the flow of energy in one direction is larger than
the flow of energy in the opposite direction. This situation is typical
for interstellar medium with its localized sources of energy. CLV02a
speculated that imbalanced turbulence can propagate over larger distances
and feed energy to clouds without star formation. To what degree
does this process get
modified in the presence of compressibility? Alfvenic modes in
imbalanced turbulence live
longer and the interaction between density fluctuations and 
the Alfven mode becomes more important.

In the field of observational studies, the situation looks remarkably 
promising. With high resolution surveys available (see Falgarone et al.
2000),
studies of turbulence statistics should at last become a mainstream
research. The prospects of studies of turbulent velocity are most
encouraging. Indeed, for the first time, one has understanding
of how to get statistics of velocity and density from channel maps
(LP00, LP03), and when the velocity centroids reflect
the statistics of velocity (LE03). Moreover, a quantitative description
of the statistics of the spectral line data cubes (see LP00, LP03) 
allows us to
devise new techniques of turbulence studies. Apart from testing of 
the particular
scaling laws, this research should identify sources and injection
scales of the turbulence (see the companion review by Falgarone et al. 1998).
 Is turbulence in molecular clouds a part
of a large scale ISM cascade (see Armstrong et al. 1997)? 
How does the share of the energy within
compressible versus incompressible motions vary within the Galactic
disk? There are examples of questions that can be answered in
future.

While this review deals mostly with power spectra, higher order statistics
will be widely used in the future. 
Recent numerical research that employed higher order
statistics (Muller \& Biskamp 2000, CLV02a, Boldyrev et al. 2002,
Cho, Lazarian \& Vishniac 2003b) showed it to be a promising tool. 
For instance, the distinction between 
the old Iroshnikov-Kraichnan and the GS95 model is difficult
to catch using power spectra with a limited inertial range, but is quite
apparent for fourth order statistics. The difference in physical 
consequences of whether the turbulence dissipates in shocks
or in intermittent vortices may be very substantial. Our discussion
in \S2 suggests that incompressible motions do dissipate via vortices.
The scaling of their intermittency with the Reynolds number
is still to be established. In the meantime, obtaining
higher order statistics
with spectral line observations is a challenging
problem. Higher order statistics obtained
from observational data 
were reported for observed velocity in Falgarone et al.
(1994)  and for density  in Padoan et al. 2003. 
According to Falgarone \& Puget (1995) and
Falgarone et al. (1995) (see also review by Falgarone),
the intermittency in vorticity distribution
can result in the outbursts of localized dissipation that
make tiny regions within cold diffuse clouds chemically active.
The testing of those ideas is done in Pety \& Falgarone
(2000), where synthetic maps obtained using
{\it hydrodynamic} simulations were analyzed. Since then,
more results indicating the hydrodynamic simulations
may to some degree reflect the physics of MHD turbulence have emerged. First
of all, CLV02a found out that the fluid motions perpendicular
to ${\bf B}$ are {\it identical} to hydrodynamic motions. 
Moreover, for low ionization the turbulence
in neutrals and ions decouple with turbulence in neutrals
forming a hydrodynamic cascade (see \S5).


Studies of the statistics of magnetic field in molecular clouds is
the next challenging problem (see Crutcher, Heiles \& Troland 2003, 
Ostriker 2003). 
Better understanding of grain alignment
(see review by Lazarian 2003 and references therein) allows to better identify variations of
polarization with the variations of magnetic field. However, this important
research has not gained sufficient momentum yet.

\section{Summary}
~~~~1. Understanding of molecular clouds requires understanding of the basics
of MHD turbulence.
MHD turbulence is not a mess. Scaling relations for its modes have been
established recently.

2. Fast decay of MHD turbulence is not due to strong coupling of
compressible and incompressible motions. The transfer of
energy from Alfven to compressible modes is small. The Alfven mode
develops on its own and decays fast.

3. Doppler shifts imprinted in spectra lines
provide an excellent way of testing theoretical expectations.
Advances in understanding how this information can be
extracted from spectroscopic data allow this.\\
{\bf Acknowledgments}{\it 
A.L. acknowledges  the support of NSF Grant AST-0125544}.

\end{article}
\end{document}